\documentclass[fleqn,usenatbib]{mnras}

\usepackage{newtxtext,newtxmath}

\usepackage[T1]{fontenc}

\usepackage{graphicx}	
\usepackage{amsmath}	
\usepackage{amssymb}	

\title[SIDM induced delay of SMBH growth]{Self-Interacting Dark Matter and the Delay of Super-Massive Black Hole Growth}

\author[A. Cruz et al.]{
A. Cruz,$^{1}$\thanks{E-mail: admcruz@uw.edu} \thanks{National Science Foundation Graduate Research Fellow}
A. Pontzen,$^{2}$
M. Volonteri,$^{3}$
T. R. Quinn,$^{4}$
M. Tremmel,$^{5}$
A. M. Brooks,$^{6}$
\newauthor
N. N. Sanchez,$^{4}$
F. Munshi,$^{7}$
and A. Di Cintio$^{8}$$^{, 9}$
\\
$^{1}$Department of Physics, University of Washington, Seattle, Washington, USA\\
$^{2}$Department of Physics and Astronomy, University College London, 132 Hampstead Road, London, NW1 2PS, United Kingdom \\
$^{3}$Institut d'astrophysique de Paris, 98bis Boulevard Arago, 75014 Paris, France\\
$^{4}$Department of Astronomy, University of Washington, Seattle, Washington, USA\\
$^{5}$Yale Center for Astronomy \& Astrophysics, Physics Department, P.O. Box 208120, New Have, CT 06520, USA\\
$^{6}$Department of Physics \& Astronomy, Rutgers,The State University of New Jersey,136 Frelinghuysen Rd. Piscataway, NJ 08854, USA\\
$^{7}$ Department of Physics \& Astronomy, University of Oklahoma, 440 W. Brooks St., Norman, OK 73019\\ 
$^{8}$Instituto de Astrof\'isica de Canarias, Calle Via L\'actea s/n, E-38206 La Laguna, Tenerife, Spain\\
$^{9}$Universidad de La Laguna, Avda. Astrof\'isico Fco. S\'anchez, La Laguna, Tenerife, Spain\\ 
}

\date{Accepted XXX. Received YYY; in original form ZZZ}

\pubyear{2020}

\begin{document}
\label{firstpage}
\pagerange{\pageref{firstpage}--\pageref{lastpage}}
\maketitle

\begin{abstract}
Using cosmological hydrodynamic simulations with physically motivated models of super-massive black hole (SMBH) formation and growth, we compare the assembly of Milky Way-mass ( $M_{\mathrm{vir}} \approx 7 \times 10^{11}$ $M_{\odot}$ at $z = 0$) galaxies in cold dark matter (CDM) and self-interacting dark matter (SIDM) models. Our SIDM model adopts a constant cross-section of 1 cm$^2$/g. We find that SMBH formation is suppressed in the early universe due to SIDM coring. SMBH-SMBH mergers are also suppressed in SIDM as a consequence of the lower number of SMBHs formed. Lack of initial merger-driven SMBH growth in turn delays SMBH growth by billions of years in SIDM compared to CDM. Further, we find that this delayed growth suppresses SMBH accretion in the largest progenitors of the main SIDM galaxies during the first 5 Gyrs of their evolution.  Nonetheless, by $z = 0.8$ the CDM and SIDM SMBH masses differ only by around 0.2 dex, so that both remain compatible with the $M_{BH}-M_{*}$ relation.  We show that the reduced accretion causes the SIDM SMBHs to less aggressively regulate star formation in their host galaxies than their CDM counterparts, resulting in a factor of 3 or more stars being produced over the lifetime of the SIDM galaxies compared to the CDM galaxies. Our results highlight a new way in which SIDM can affect the growth and merger history of SMBHs and ultimately give rise to very different galaxy evolution compared to the classic CDM model. 
\end{abstract}




\section{Introduction} \label{intro}

The favoured model for dark matter (DM) has long been cold DM (CDM), a single, collision-less particle species with negligible primordial thermal dispersion. Although DM-only simulations of $\Lambda$CDM cosmology prove to be very successful on large scales, they predict high central DM density "cusps" \citep{Navarro1997,Moore1998,Bullock2001,Wechsler2002}, directly in conflict with observations of dwarf irregulars \citep{Flores1994,Moore1994,DeBlok1997}, low surface brightness galaxies \citep{deBlok2001,Kuzio2008,Kuzio2011}, nearby field dwarfs \citep{deBlok2008,Oh2008}, low-mass spiral galaxies \citep{Gentile2004,Simon2005,Castignani2012,Adams2014}, and satellite dwarfs of the Milky Way (MW) \citep{Walker2011,Salucci2012,Breddels2013}. These observations suggest the existence of kpc-scale DM cores, a discrepancy with $\Lambda$CDM known as the core-cusp problem.
\citet{Spergel2000} were among the first to point out that dark matter self-interactions with a mean free path ranging from 1 kpc to 1 Mpc would naturally impact galaxies and clusters of galaxies, while simultaneously preserving the large scale success of $\Lambda$CDM. 
Thus, Self-interacting DM (SIDM) was proposed to address the core-cusp problem over a decade ago \citep{Spergel2000,Burkert2000} as it naturally produces cored profiles in the inner 1~kpc of DM haloes in DM-only simulation,  which better agree with observations. 

\hfill \break
\indent On the other hand, it has since been shown that baryonic physics can also create DM cores. In particular, a number of studies show that outflows driven by supernovae (SNe) cause the formation of shallow DM density profiles at the centers of galaxies \citep{Read2005, Governato2012,Pontzen2012,Dicintio2014b,Dicinto2014a,ABrooks2014,Pontzen2014,Onorbe2015,Tollet2016, Benitez2019}. Moreover, high resolution cosmological simulations which include baryonic physics produce indistinguishable DM distributions in the central regions of dwarfs, which agree well with observations, in CDM and SIDM models \citep[][hereafter BF15]{BastidasFry2015}. This indicates that 1) SIDM and CDM both remain viable DM candidates and 2) baryonic physics in simulations is required to make reliable predictions about the nature of DM. 
\hfill \break 
\indent Despite the successes of simulations of CDM with baryons, they fail to simultaneously produce the densest galaxies observed \citep{Santos2018}. This seems to be true even for the highest resolution simulations which should be able to reproduce the compactness observed in some galaxies \citep{GK2019}. Meanwhile, a series of works have demonstrated that a SIDM model with an interaction cross-section of $\sim 3$ cm$^2$/g can reproduce galaxy rotation curves from $\sim 50$ to 300 km/s \citep{Kamada2017,Ren2019,kaplinghat2019dark}. SIDM also produces a trend in central density of MW-satellites as a function of orbital pericenter \citep{Nishikawa2019,Kahlhoefer2019} which agrees extremely well with Gaia data \citep{Kaplinghat2019} and has not been found in CDM simulations with baryons. These results emphasize that SIDM is becoming an increasingly interesting DM candidate. 

However, few cosmological simulations with baryons and SIDM exist to date with most work focusing on dwarf galaxies given the tensions with observations at this mass scale. \citet{VogelsbergerZavala2016} examined MW-mass simulations of SIDM with a single DM particle which was allowed to interact with itself and with a massless neutrino-like fermion (dark radiation) but did not include super-massive black holes (SMBHs). More recently \citet{DiCintio2017b} compared MW-mass galaxies, as well as lower mass galaxies, from cosmological volumes with side lengths of 8 Mpc in CDM and SIDM with an interaction cross-section of 10 cm$^2$/g and found that at low-$z$ the most massive SIDM SMBHs in galaxies were routinely off-centre from their hosts unlike the CDM SMBHs which remained at their hosts centre. 

 In this paper we expand on the work of \citet{DiCintio2017b} by using simulations with baryonic physics and physically motivated models of SMBH formation and growth through mergers \citep[][hereafter T15]{Tremmel2015} and accretion \citep[][hereafter T17]{Tremmel2017} to study the effects SIDM has on the formation and evolution of SMBHs and their  MW-mass host galaxies. MW-mass galaxies are a particularly interesting place to examine SIDM's effect on SMBH growth because these galaxies may or may not be quenched and straddle the regime where SMBHs start to suppress star formation.  Unlike \citet{DiCintio2017b} we used a much lower SIDM cross-section of 1 cm$^2$/g and examined SMBH formation and the temporal evolution of SMBHs and their host galaxies in CDM and SIDM cosmologies, starting from high-$z$. This is important given that the early universe is much denser than today, and thus the SIDM interaction rate peaks at high-$z$ \citep{Roberston2015}. Still more, in the early universe the effects of coring due to bursty SNe feedback detailed in BF15 are not present for the galaxies we considered given that they have stellar masses less-than or equal to $10^6 M_{\odot}$, the threshold for star formation to start to core DM haloes \citep[see Figure 7 of][]{Governato2015}. Thus the distinct coring effects of SIDM are preserved.  

The organization of this paper is as follows: in Section \ref{simulations} we will detail the physics integrated into our suite of simulations. Section \ref{results} will detail our results with a discussion of differences in CDM vs. SIDM cosmologies, including star formation and gas content \ref{subsec:SFandGas}, SMBH formation \ref{subsec:BHForm} and eventual growth through mergers and accretion \ref{subsec:BHgrowth}. In section \ref{sec:discussion} we will discuss the implications of our study as well as summarize and conclude the paper. 

\section{Simulations} \label{simulations}

The simulations examined in this paper were run in a fully cosmological context to $z = 0$. However, this paper will focus on the evolution from $z = 20$ to $z \approx 0.8$, as this time interval is where we see the biggest differences in SMBH growth and star formation in the CDM vs. SIDM runs. Furthermore all simulated zoom-in galaxies quench near or at z $\approx$ 0.8, and remain quiescent for billions of years. Further, all galaxies have formed $\gtrapprox$ 90$\%$ of their total stellar mass by z $\approx$ 0.8. 

The simulations were run using Charm N-body GrAvity Solver (ChaNGa\footnote{www-hpcc.astro.washington.edu/tools/changa.html}), a smoothed particle hydrodynamics (SPH) N-body tree code \citep{Menon2015}. ChaNGa is a successor of GASOLINE and thus includes the same models for low temperature metal line cooling, self shielding, star formation \citep[using a][IMF]{Kroupa2001}, "blastwave" SNe feedback, and cosmic UV background \citep{Wadsley2004,Wadsley2008,stinson2006}. The SPH implementation also includes thermal diffusion \citep{Shen2010} and eliminates artificial gas surface tension by using a geometric mean density in the SPH force expression \citep{Ritchie2001,Menon2015,Governato2015,Wadsley2017}. This addition better simulates shearing flows with Kelvin-Helmholtz instabilities.

In all simulations we assumed a $\Lambda$ dominated cosmology \citep[$\Omega_{\mathrm{m}} = 0.3086$, $\Omega_{\Lambda} = 0.6914$, $h = 0.67$, $\sigma_8 = 0.77$,][]{Plank2014} and used the ``zoom-in" methods described by \citet{Pontzen2008}. All simulations are run with a Plummer equivalent softening length, $\epsilon = 250$~pc and mass resolution of 1.4 $\times 10^5 $ M$_{\odot}$ and 2.1 $\times 10^5 $ M$_{\odot}$ for DM and gas particles, respectively. We used two DM models: the standard CDM model and an SIDM model with a constant interaction cross-section of $\sigma_{\mathrm{dm}} = 1$ cm$^2$/g. We re-simulate two of the MW-mass galaxies ($M_{\mathrm{vir}} \approx 7 \times10^{11} M_{\odot}$ at $z = 0$, see Equation \eqref{eqn:Mvir}) presented in \citet{Sanchez2019} with SIDM (their GM2 and GM3 galaxies). As the GM moniker suggests, these galaxies are part of a series constructed using ``genetic modification'' \citep{Roth2016}, in the case of GM2 and GM3 this allows us to robustly test the effects of SIDM on SMBH formation and subsequently, star formation. Specifically, we simulated GM3 (which has quenched star formation) in order to examine strong effects of SMBHs, which play a large role in quenching galaxies \citep{Pontzen2017}. GM2 (also quenched) allows us to check for sensitivity to small changes. The GM runs considered have almost identical histories but differ due to an altered satellite population, which results in a different rate of accretion at early times; for more information see \citet{Sanchez2019}. Our use of GM2 and GM3 allows us to specifically verify that our SIDM results are robust to this aspect of the assembly of a halo. In this paper, we use the same nomenclature as \citet{Sanchez2019} for the CDM galaxies, i.e, GM2, GM3 and append `SI1' (i.e. GM3SI1, etc.) for the counterpart simulations run with SIDM with an interaction cross-section of $\sigma_{\mathrm{dm}} = 1$ cm$^2$/g. 

After running our simulations we extract all of our main haloes and sub-haloes using the AMIGA halo finder \citep{Knollmann2009}. We calculate the virial mass of haloes as: 
\begin{equation}\label{eqn:Mvir}
    M_{\mathrm{vir}} = \frac{4}{3}\pi\Delta_{h} \overline{\rho} R^3_{\mathrm{vir}}
\end{equation}

\noindent where $\overline{\rho}$ is the critical density of the Universe, $\Delta_{\mathrm{h}} = 200$ is the over-density threshold, and $R_{\mathrm{vir}}$ is the halo virial radius.  

\subsection{Self-interacting Dark Matter Physics} \label{subsec:SIsec}

The SIDM implementation used in our simulations closely follows the standard Monte Carlo method described in detail in BF15. We briefly describe the features of this model here and refer the reader to BF15 and references therein for details. SIDM interactions are modeled under the assumption that each simulated DM particle represents a
patch of DM phase-space density and that the probability of collisions is derived from the collision term in the Boltzmann equation.
Collisions are elastic and explicitly conserve energy and momentum.  When a particle collision is detected, particles are isotropically and
elastically scattered to random angles. For a detailed discussion see also \citet{Rocha2013} and
\citet{Yoshida2000,Donghia2003,Kaplinghat2014}. The SIDM interaction rate of particles will vary with local DM density $\rho_{\mathrm{dm}}(r, z)$, cross-section $\sigma_{\mathrm{dm}}$ and velocity dispersion $v(r,z)$ as

\begin{equation} \label{eqn:SIrate}
\Gamma_{\mathrm{SI}}(r,z) \backsimeq \rho_{\mathrm{dm}}(r,z) v(r,z) \sigma_{\mathrm{dm}}
\end{equation}

\noindent up to an $\mathcal{O}(1)$ constant \citep{Rocha2013}. Collisions between SIDM particles result in energy exchange, which heat the halo centre until it becomes isothermal \citep{Balberg2002,Colin2002,Koda2011}.

The SIDM cross-section $\sigma_{\mathrm{dm}}$ must adhere to several astrophysical observations, including the necessity of forming DM cores in faint galaxies without the over-evaporation of MW-mass galaxy satellites or galaxies in clusters and maintaining the elliptical shape of haloes and clusters \citep{Firmani2001,Gnedin2001,Peter2013,Robertson2017b}. Utilizing SIDM-only simulations with these observations in mind, authors have found the relevant range to impact galaxy evolution and avoid upper limits to be $0.1$ cm$^2$/g $< \sigma_{\mathrm{dm}} < 1 $ cm$^2$/g for velocity-independent cross-sections \citep{Vogelsberger2012,Peter2013,Rocha2013,Vogelsberger2013,Zavala2013,VogelsbergerZavala2016,Cyr2016}. A velocity-dependent cross-section could ease the constraints on $\sigma_{\mathrm{dm}}$ by allowing DM to behave as a collisional fluid on the scale of dwarfs, and more collision-less at the scale of clusters \citep{Yoshida2000,Colin2002,Elbert2018}. Velocity-dependent cross-sections can also influence when the SIDM interaction rate peaks as a function of redshift \citep{Roberston2015}. Further, a velocity-dependent cross-section allows for cross-section of 3 cm$^2$/g for range of rotational velocities explored in \citet{Ren2019} and \citet{kaplinghat2019dark}.  The interaction cross-section $\sigma_{\mathrm{dm}}$ for all SIDM runs in this work was set to 1 cm$^2$/g.

\subsection{Star Formation} 
\label{sf_subsec}
Since this paper is largely focused on star formation, we review here in more detail the star formation prescription \citep{stinson2006} and parameters (T17) used in our simulations. 

Gas particles are allowed to form stars in our simulations if they surpass minimum density ($n_*$) and maximum temperature ($T_*$) thresholds. The probability of creating a star particle from gas with dynamical time $t_{\mathrm{dyn}}$ and characteristic star formation time, $\Delta t$, assumed to be $10^6$~years is given as:

\begin{equation} \label{eqn:starform}
p = \frac{m_{\mathrm{gas}}}{m_{\mathrm{star}}} (1-\mathrm{e}^{c_* \Delta t/t_{\mathrm{dyn}}})
\end{equation}

\noindent where $c_*$ is the star formation efficiency. Further, star formation is regulated by the fraction of SNe energy that is coupled to the ISM, our star formation efficiency ($c_*$), and our density and temperature thresholds. The values we have adopted for our sub-grid parameters are as follows:

\begin{description}
 \item[$\bullet$] star formation efficiency $c_{*} = 0.15$
 \item[$\bullet$] Gas density threshold, $n_{*} = 0.2$ cm$^{-3}$ 
 \item[$\bullet$] Gas temperature threshold, $T_{*} = 10^4$ K 
 \item[$\bullet$] SNe energy coupling efficiency, $\epsilon_{\mathrm{SN}}$ of 75 percent 
\end{description}

SNe feedback adopts a `blastwave' implementation \citep{stinson2006} and gas cooling is regulated by metal abundance as in \citet{Guedes2011}. 

\subsection{Black Hole Physics}
\label{subsec:BHphys} Our simulations also include SMBH formation and improved SMBH accretion and feedback models which explicitly follows the orbital evolution of SMBHs (T15; T17). 

The SMBH seed (with seed mass of 10$^6$ M$_{\odot}$) formation is connected to the physical state of the gas in the simulation at high-$z$, without assumptions about the halo occupation fraction. SMBHs seeds form in the early universe if the gas particle has already met the star formation thresholds (see Subsection \ref{sf_subsec}) and additionally has: 

\begin{description}
 \item[$\bullet$] Low metallicity ( $Z   < 3 \times 10^{-4}$)
 \item[$\bullet$] Density 15 times that of the star formation threshold (3 cm$^{-3}$) 
 \item[$\bullet$] Temperature between 9500 and 10000 K 
 
\end{description}

This seeding method allows SMBHs to naturally populate galaxies of different masses with SMBHs. Seed SMBH formation is limited to the highest density peaks in the early universe with high Jeans masses and to gas that is cooling relatively slowly, thus approximating SMBH formation sites consistent with those predicted for SMBH seed formation \citep{Volonteri2012}. This seeding method forms most SMBH seeds within the first Gyr of the simulation, which allows us to follow SMBH dynamics throughout the assembly of the host halo, even for small haloes. 

Another important improvement in the SMBH model utilized in these simulations is the treatment of dynamical friction, the gravitational wake of a massive body moving in the extended potential of a medium, which will cause the orbit of SMBHs to decay towards the centre of massive galaxies \citep{Chandrasekhar1943, BT2008}. Previously, authors have used analytic expressions to compute the dynamical friction timescale t$_{\mathrm{df}}$ of rigid bodies merging in the centre of galaxies \citep[i.e.][]{Taffoni2003,Boylan2008}, demonstrating that this timescale can easily exceed several Gyrs \citep{DiMatteo2005,Sijacki2007, Schaye2015}. The advection technique repositions and forces SMBHs to the galaxy centre during merger events or during satellite accretion, and therefore lacks realistic sinking timescales for SMBHs within galaxies. In this work we instead use a prescription of dynamical friction which explicitly follows the orbital evolution of SMBHs, introduced by T15. This prescription utilizes a sub-grid approach for modeling unresolved dynamical friction on scales smaller than our gravitational softening length, adding a force correction to the SMBH acceleration. The SMBH then experiences a dynamical friction force according to: 

\begin{equation} \label{eqn:dynamF}
    \textbf{F}_{\mathrm{df}} = - 4\pi G^2 M_{\mathrm{BH}} \rho_{\mathrm{host}}(< v_{\mathrm{BH}})\ln{(\Lambda)}\frac{\textbf{v}_{\mathrm{BH}}}{v^3_{\mathrm{BH}}} 
\end{equation}

\noindent where $M_{\mathrm{BH}}$ is the mass of the SMBH, $v_{\mathrm{BH}}$ is the speed of the SMBH relative to the local centre of mass velocity, $\rho_{\mathrm{host}}( < v_{\mathrm{BH}})$ is the density of the host background particles with velocities less than the $v_{\mathrm{BH}}$. The Coulomb logarithm, ln($\Lambda$), depends on the minimum and maximum impact parameters such that ln($\Lambda$) $\sim$ ln($b_{\mathrm{max}}/b_{\mathrm{min}}$). Given that dynamical friction is well resolved at scales above a softening length we take the maximum impact parameter, $b_{\mathrm{max}} \sim \epsilon$ to avoid double counting. The minimum impact parameter, $b_{\mathrm{min}}$, is taken to be the minimum 90$\degr$ deflection radius with a lower limit set to be the Schwarzschild radius (see T15 for more details). This acceleration from Equation \eqref{eqn:dynamF} is added to the SMBH's current acceleration and integrated in the following time step. The resulting sinking timescale $t_{\mathrm{df}}$ will thus be dependent on the density of the surrounding galaxy, and on the mass and velocity of the SMBH itself. T15 showed that this technique produces realistically sinking SMBHs. Correctly accounting for this timescale can lead to SMBH pairs that exist at kpc-scale separations for several Gyrs \citep{Tremmel2018a}. In CDM simulations, it is possible to have ``wandering" SMBHs with sinking timescales longer than a Hubble time \citep{Bellovary2010, Bellovary2019, Tremmel2018b}, an effect that can be exacerbated by the lower central densities caused by SIDM \citep{DiCintio2017b}.

The SMBHs in our simulation also obey a modified Bondi-Hoyle accretion which accounts for the rotational support of the surrounding gas. Our SMBHs accrete according to T17: 
\begin{equation} \label{eqn:Mdot}
\begin{split}
    \dot{M}_{\mathrm{BH}} &= \alpha\pi(GM_{\mathrm{BH}})^2\rho \times   \left\{
\begin{array}{lll}
      \dfrac{1}{(v^2_{\mathrm{bulk}} + c^2_{\mathrm{s}})^{3/2}} & v_{\mathrm{bulk}} > v_{\theta}\\
     \\
      \dfrac{c_s}{(v^2_{\theta} + c^2_s)^2} & v_{\mathrm{bulk}} < v_{\theta} \\
\end{array} 
\right. 
\\
    \alpha &= \left\{
	\begin{array}{ll}
		\Big(\frac{n}{n_{*}}\Big)^{\beta} & n \geq n_{*} \\
		\\
		\ 1 &   n < n_{*}
	\end{array} 
	\right.
\end{split}
\end{equation}

\noindent where $\rho$ is the local gas density, and $c_{\mathrm{s}}$ is the sound speed of the gas. Values for density and temperature of nearby gas are estimated from smoothing over the 32 nearest gas particles and accretion is not allowed to occur from gas particles farther than 4 $ \times \ \epsilon$. The tangential velocity, $v_{\theta}$, is derived from the resolved kinematics of nearby gas particles and compared to $v_{\mathrm{bulk}}$, the overall bulk motion of the gas. When either the bulk motion or internal energy of the gas dominates over rotational motion, the accretion model converges to the  Bondi-Hoyle prescription. In either case, we add a boost factor, $\alpha$, calculated by comparing the density of nearby gas particles to our star formation density threshold. This is considered to be the threshold beyond which our simulation no longer fully resolves the internal structure of gas. For lower densities, we assume that the gas is not sufficiently multiphase to require such a boost, as in \citet{Booth2009}. How much this boost increases with density is governed by $\beta$, a free parameter, which is set to 2, as is discussed in section 5.5 of T17. This accretion prescription will thus naturally limit accretion of SMBHs that form in unfavorable environments such as in dwarf galaxies. 

Energy from accretion is then isotropically transferred to nearby gas particles with a technique similar to the blast wave SNe feedback of \citet{stinson2006}; i.e., gas cooling is turned off for the gas particles immediately surrounding the SMBH, which resembles the continuous transfer of energy during each SMBH time step. This cooling shut off is only for a single BH time step (typically 10$^4$-10$^5$ years) and has been shown to result in large scale outflows that can quench star formation in massive galaxies and enrich the circumgalactic medium \citep{Pontzen2017,Tremmel2019,Sanchez2019}. The rate at which energy is coupled to the surrounding gas particles is given by, 

\begin{equation}\label{eqn:BHinj}
	\dot{E}_{\mathrm{BH}} = \epsilon_{\mathrm{r}} \epsilon_{\mathrm{f}}\dot{M}_{\mathrm{BH}}c^2
\end{equation}

\noindent where $\dot{M}_{\mathrm{BH}}$ is the accretion rate defined in Equation \eqref{eqn:Mdot} and $\epsilon_{\mathrm{r}} = 0.1 $ and $\epsilon_{\mathrm{f}} = 0.02$ are the radiative and feedback efficiencies, respectively and $c$ is the speed of light.

It should be noted that the sub-grid parameters used in this work to regulate star formation and feedback from SMBHs and SNe were optimized against a comprehensive set of $z = 0$ galaxy scaling relations using a multi-dimensional parameter search as detailed in T17. It is important to note that the sub-grid physics have in no way been optimized to produce any characteristics at high-$z$, making all of the high-$z$ evolution of SMBHs in these simulations purely predictions of the simulation.

\section{Results} \label{results}

In this section, we examine the star formation and gas content of CDM vs. SIDM simulations, and how gas content is connected to the activity of SMBH and SNe feedback. The formation of SMBHs in CDM and SIDM is considered, as well as the growth of the central SMBH in the largest progenitors of what become the main MW-mass galaxies at $z = 0$. We detail the growth of the central SMBH through mergers and through accretion. Differences in SMBH feedback are then examined along with the connection between SMBH feedback and star formation.

\subsection{Star Formation and Gas Content} \label{subsec:SFandGas}

\begin{figure}
\includegraphics[width=\columnwidth]{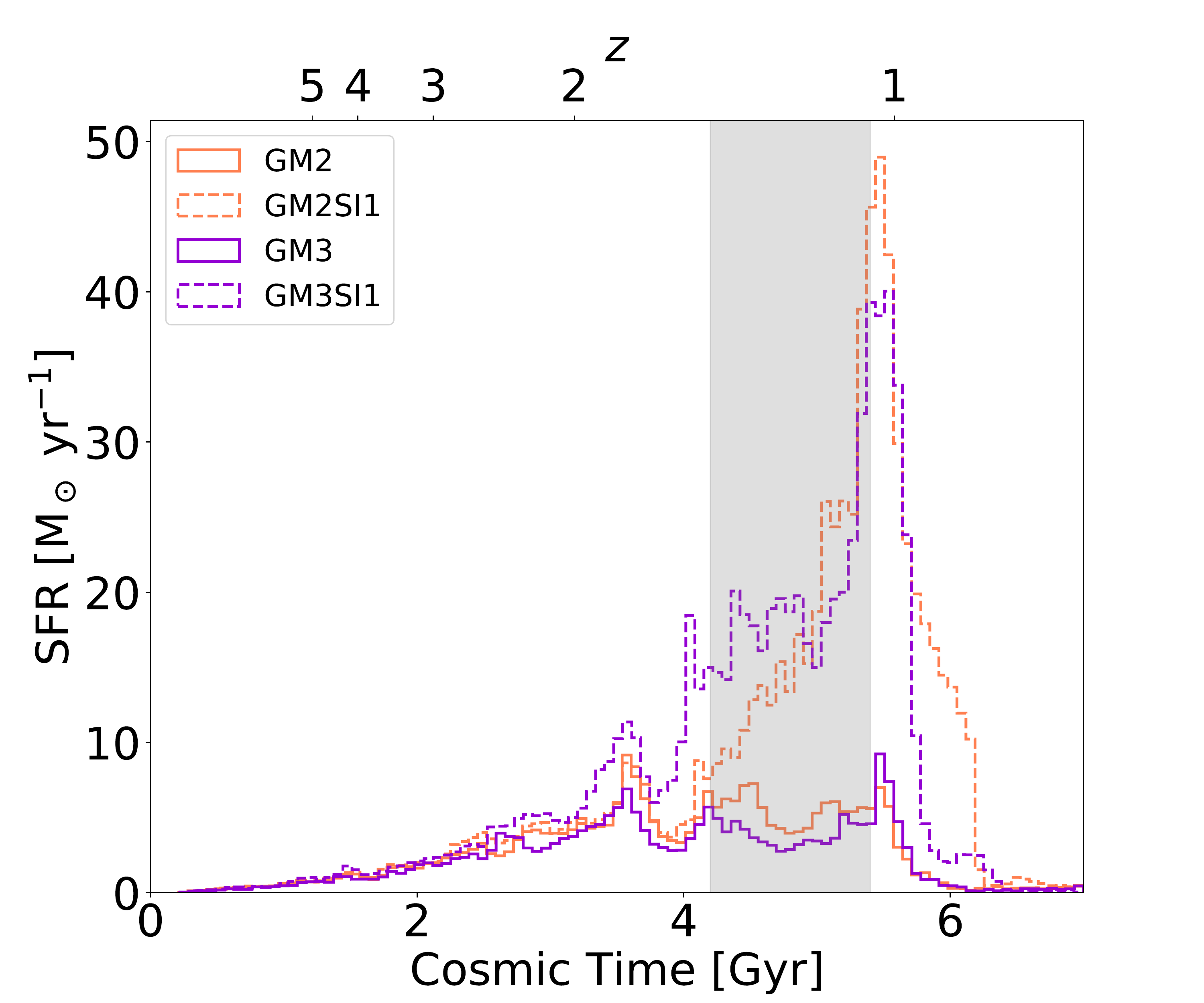}
    \caption{Star formation rate as a function of cosmic time from $t = 0 - 7$ Gyr. There is a striking difference between the GM3 and GM3SI runs, as well as in the star formation between GM2 and GM2SI1. The main halo of GM3SI1 ($M_{\mathrm{vir}} \approx 8 \times 10^{11} \mathrm{M}_{\odot}$) produces 3.7 times more stars than the CDM GM3 galaxy over its lifetime, but still quenches near 6 Gyrs. The GM2SI1 produces 3.1 times more stars compared to GM2 over its lifetime. GM2SI1 starts to quench a bit later (starting near $t = 6.2$~Gyrs) compared to GM2, GM3 and GM3SI1. The gray shaded region represents the time interval in which we examine gas content.}
    \label{fig:SFRSI1}
\end{figure}

When comparing star formation histories (SFHs), we include only the most massive progenitors of the main haloes in each simulation at $z = 0$, excluding star formation from satellites. We see stark differences between the SFHs of MW-mass galaxies that clearly depend on the assumed DM physics as illustrated in Figure~\ref{fig:SFRSI1}. The zoom-in simulations with CDM (GM2 and GM3, solid orange and purple curves, respectively) show MW-mass galaxies with relatively stable star fomration rates (SFRs) of $5 - 8 ~\mathrm{M}_{\odot} \mathrm{yr}^{-1}$ over roughly 6 Gyrs of cosmic time, whereas those zoom-in simulations that include SIDM (GM2SI1 and GM3SI1, dashed orange and purple curves, respectively) exhibit much higher SFRs and overall burstier SFHs. 

The star formation in GM3SI1 starts to deviate from CDM beyond 30$\%$ around $t = 3$~Gyr (where $t$ is cosmic time) and beyond 50$\%$ starting around $t = 4.5$~Gyr. The ratio between SFR in GM3SI1 to GM3 can be as high as 8.1 near $t \approx 5.5$~Gyr where GM3SI1 goes through a ``star burst" period before eventually quenching near $t \approx 6.5$ Gys. In total, GM3SI1 produces 3.7 times more stars over its lifetime compared to GM3.

The difference in SFH between GM2SI1 and GM2 is slightly less dramatic for the first 4 Gyrs of the galaxies' lifetimes; however GM2 and GM2SI1 start to deviate from one another after $t = 4$~Gyr faster than the galaxies in the GM3-suite runs. The ``burstiness" near $t = 5.5 $~Gyr is also more enhanced in the GM2SI1 vs. GM2 runs compared to the enhancement seen in the GM3-suite runs. The ratio between SFR in GM2SI1 and GM2 can be as high as 8.6. The GM2SI1 galaxy produced 3.1 more stars over its lifetime compared to the GM2 galaxy. Halo mass, stellar mass and other properties of our galaxies at $z \approx 0.8$ ($t = 7$ Gyr) are detailed in Table \ref{tab:haloprops}. 

\begin{figure*}
\includegraphics[width=2\columnwidth]{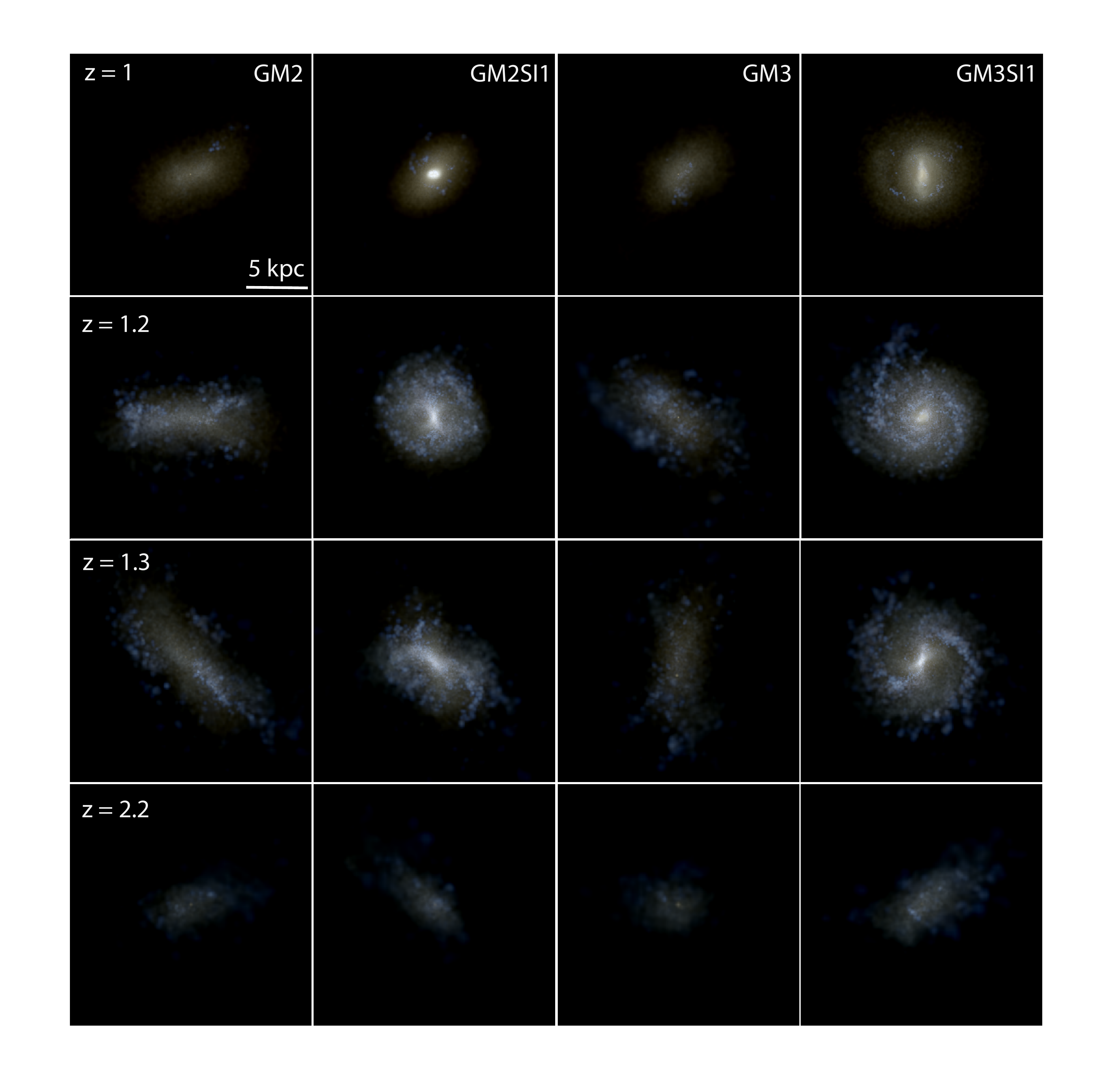}
    \caption{Cosmic evolution of stars in all four galaxies. Four snapshots taken (from the top) at $z = $ 1, 1.2, 1.3, and 2 and showing (from left to right) GM2, GM2SI1, GM3, GM3SI1.  The stars are shown in UVI colors assuming a Kroupa IMF and are oriented such that the angular momentum axis calculated from {\sc Pynbody} is in the z-direction which points out of the page. All images encompass 20 kpc on each side and the 5 kpc scale is in physical units. The surface brightness for all images ranges from 23 - 13.6 mag/arcsec$^2$.  The four redshift snapshots are chosen to sample four different ``epochs'' that are present in the SFHs of our galaxies. From bottom up, at $z~=~2.2$ GM3SI1 has started to deviate from GM3 in star formation, but GM2SI1 and GM2 remain similar, which is clear in the similar UVI images at this redshift. The $z~=~1.3$ snapshot represents an epoch where the star formation in the SIDM galaxies deviate from the CDM galaxies beyond the 50$\%$ level; at this redshift the SIDM galaxies are spiral galaxies and their CDM counterparts are irregular. The $z~=~1.2$ snapshot represents the epoch near the SIDM galaxies ``star burst'' period; here the SIDM galaxies retain their spiral morphology while the CDM galaxies remain more irregular. By $z = $ 1, the CDM galaxies are quenching and turning red; the SIDM galaxies are also turning red, but remain much brighter in their centers compared to the CDM galaxies.}
    \label{fig:uvi}
\end{figure*}

The DM-model-dependent differences seen in the SFHs are also apparent in UVI images of the stars in each galaxy. Each row of Figure \ref{fig:uvi} samples a different ``epoch" of the SFH and each column shows one of the four simulated galaxies. From bottom to top, the $z~=~2.2$ snapshot represents the first epoch, where we see differences of up to 30$\%$ in the SFHs of the SIDM galaxies compared to the CDM galaxies. During this epoch, the SIDM galaxies are only slightly brighter than the CDM galaxies. The $z~=~1.3$ snapshot samples the second epoch, where the star formation in the SIDM galaxies deviates from the CDM galaxies beyond 50$\%$. In this epoch, the CDM galaxies are irregular, whereas the SIDM galaxies are more spiral and much brighter in their centers. The $z~=~1.2$ snapshot represents the epoch just before the ``star burst" in the SIDM galaxies. Here, the SIDM galaxies retain their spiral morphology while the CDM galaxies remain irregular. Finally, the $z~=~1$ snapshot represents the epoch when the CDM galaxies quench. In the CDM galaxies the overall surface brightness decreases and the galaxies appear more elliptical whereas the SIDM galaxies surface brightness decreases in the outer regions but remains high in their centers. Thus, examining UVI images of the CDM and SIDM galaxies in each suite demonstrates DM-model-dependent differences in morphology and stellar evolution. 

\begin{figure}
\includegraphics[width=\columnwidth]{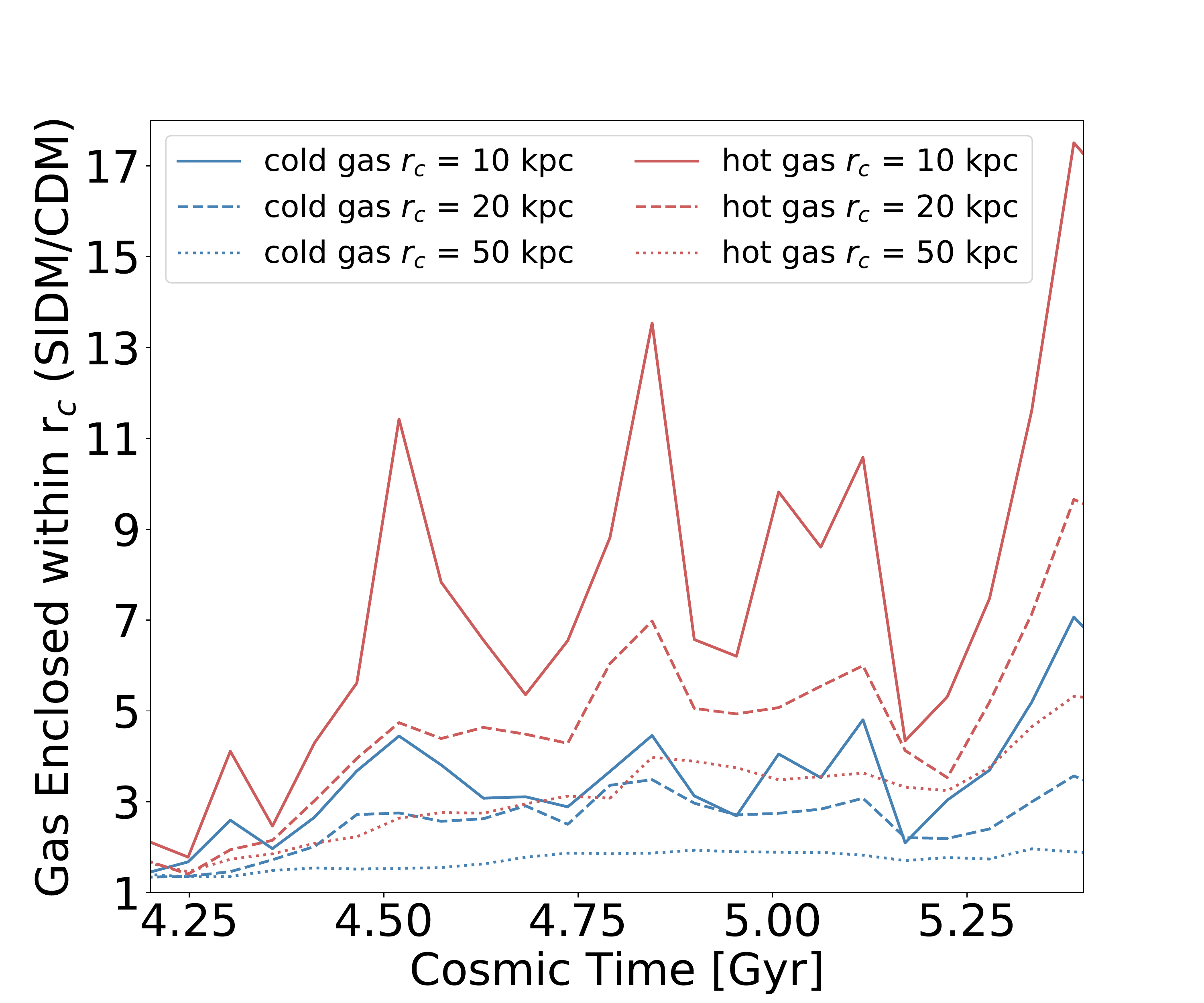} 
    \caption{Ratio of GM3SI1 gas compared to GM3 gas enclosed within a spherical volume with cut off radius of $r_{\mathrm{c}}$. Gas content is broken down into hot gas ($T > 10^5 $ K) and cold gas, where cold gas is defined to be gas that satisfies our temperature criteria for star formation ( $T < 10^4$ K). The hot gas is shown in red where as the cold gas is shown in blue. Various line styles are used to indicate gas in different cut off radii. The ratio of cold gas within the inner 10~kpc comes close to 7 just before the ``star burst" near $t \approx 5.5$ Gyr in GM3SI1. At every cut off radius the ratio of gas in GM3SI1 compared with GM3 is greater than 1 and gradually increases as $\mathrm{r_c}$ decreases. We use the GM3-suite as a representative example of both GM suites.} 
    \label{fig:gas_ratio}
\end{figure}

\begin{table}
\caption{GALAXY PROPERTIES AT \textbf{$z = 0.8$}}
\label{tab:haloprops}
\begin{tabular}{llllll}
\hline 
 Sim    & $M_{\mathrm{vir}}$$^\mathrm{a}$   & $M_{*}$$^\mathrm{b}$                   & $M_{\mathrm{g}}$$^\mathrm{c}$              & $M_{\mathrm{BH}}$$^\mathrm{d}$            &  $R_{\mathrm{vir}}$$^\mathrm{e}$\\ 
        & $M_{\odot}$          & $M_{\odot}$                & $M_{\odot}$                   & $M_{\odot}$                  & kpc \\
 \hline
 GM2    & 4.6 $\times 10^{11}$ &   1.3 $\times 10^{10}$     &   3.9 $\times 10^{10}$        &       5.1 $\times 10^7$      &           140.8    \\
 GM2SI1 & 5.1 $\times 10^{11}$ &   3.6 $\times 10^{10}$     &   6.2 $\times 10^{10}$        &       4.6 $\times 10^7$      &           148.6    \\
 GM3    & 4.5 $\times 10^{11}$ &   9.8 $\times 10^{9}$      &   3.5 $\times 10^{10}$        &       6.2 $\times 10^7$      &           139.2     \\
 GM3SI1 & 5.0 $\times 10^{11}$ &   3.6 $\times 10^{10}$     &   5.2 $\times 10^{10}$        &       4.0 $\times 10^7$      &           147.5     \\ \hline
\end{tabular}

\footnotesize{$^\mathrm{a}$ Halo virial mass as defined in Equation \eqref{eqn:Mvir}.} \\
\footnotesize{$^\mathrm{b}$ Total stellar mass within $R_{\mathrm{vir}}$.} \\
\footnotesize{$^\mathrm{c}$ Total gas mass within $R_{\mathrm{vir}}$.} \\
\footnotesize{$^\mathrm{d}$ Mass of central SMBH in major progenitor of MW-mass galaxies.}\\
\footnotesize{$^\mathrm{e}$ Halo virial radius, $R_{\mathrm{vir}}$.}
\end{table}

\begin{figure}
\centering 
\includegraphics[width=\columnwidth]{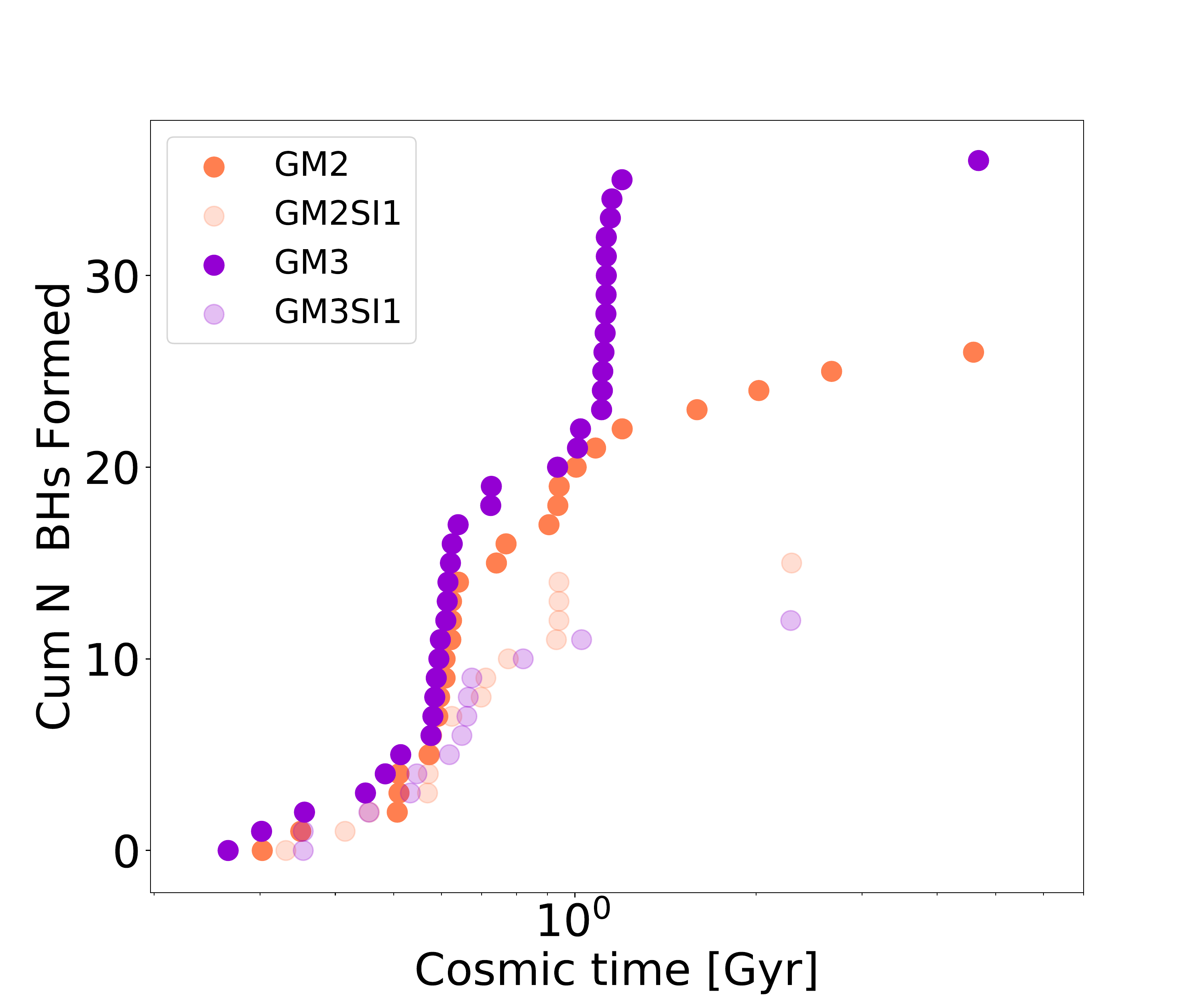}
    \caption{Cumulative number of SMBH seeds vs. log of cosmic time. In both the GM2-galaxy and the GM3-galaxy suites the CDM galaxies produce more SMBHs in the first 2 Gyrs of the simulation. GM3 produces 2.9 times more SMBHs that GM3SI1, where as GM2 produces 1.7 times more SMBHs than GM2SI1.}
    \label{fig:BHhist}
\end{figure}

These differences are substantial given that the two sets of simulations come from the exact same ICs, respectively, with only the underlying DM models changing. While \citet{BKeller2019} has emphasized that star formation rates can vary stochastically between different runs due to purely numerical artefacts, this is not the cause of our differences here. First, the extent of our differences are vastly higher than those found by \citet{BKeller2019} and, second, the identical trends seen when switching to SIDM in GM2 and GM3 serves as an independent robustness check.  The large difference we see in SFH must be attributable to changes in the properties of the galaxy's gas. Excess gas that is dense ($n \geq$ 0.2 $m_{\mathrm{p}} / $ cm$^3$) and cool ($T \leq$ 10$^4$ K) can turn into stars in our simulations. Thus, we next examine the gas content of the SI1 galaxies compared to the CDM galaxies.

In Figure~\ref{fig:gas_ratio} we examine the gas content within spheres of various cutoff radii, $r_{\mathrm{c}}$, centered on the shrinking sphere centre (found using methods of \citet{Power2003} implemented in \citet{Pontzen2013}) of the most massive progenitors of the main galaxy at $z = 0$ and compare the mass in gas between SIDM and CDM in the GM3-suite. The gas is broken down into hot gas and cold gas, where cold gas is the gas that satisfies our temperature criteria for star formation and hot gas is gas with $T > 10^5$ K. Figure~\ref{fig:gas_ratio} shows cosmic time between 4.2 Gyrs and 5.4 Gyrs, leading up to the "star burst", or the peak of star formation, in SIDM and after both the CDM and SIDM galaxies undergo mergers. The period of time for which we examine the gas properties is highlighted in gray in Figure~\ref{fig:SFRSI1}. The SFH between the CDM and SIDM galaxies begins to deviate beyond 30$\%$ around $t = 3$~Gyrs and beyond 50$\%$ after $t\approx$ 4 Gyrs. The deviation accelerates rapidly after this. Thus, this is the time region that is interesting to examine in order to understand the differences in the SFH between the CDM and SIDM galaxies. The ratio of cold and hot gas decreases smoothly as $r_\mathrm{c}$ increases, indicating that there is more gas in the inner regions of the SIDM galaxy compared to the CDM galaxy.

In the inner 10~kpc SIDM has much more cold and hot gas compared to CDM. The excess cold gas is responsible for the increased number of stars formed in the SIDM galaxies. The increased number of stars formed in the SIDM galaxies are responsible for the excess hot gas, via increased SNe-feedback (which is proportional to the number of stars formed). Therefore, the excess hot gas in the SIDM galaxies is a consequence of their increased SFRs. The excess cold gas can be caused by either stronger inflows to the centre of the SIDM galaxies or cold gas being pushed out and depleted from the inner 10 kpc of CDM galaxies. Lagrangian particle tracking which matches and traces gas particles from the inner 10~kpc of the CDM and SIDM galaxies shows that in CDM galaxies gas that is diffuse at later times was more dense and structured at earlier times. This indicates that gas is being more readily disrupted in the central region in our CDM runs. We show in the next section that this is due to differences in SMBH feedback between the CDM and SIDM simulations. 

\subsection{Black Hole Formation} \label{subsec:BHForm}

In Figure~\ref{fig:BHhist}, we explore the total number of SMBHs formed in our simulations vs. cosmic time. SMBHs formation is suppressed in the SIDM runs, when compared to their CDM counterparts. GM3 forms 2.9 times as many SMBHs in the early universe compared to GM3SI1 and GM2 forms 1.7 times as many SMBHs as GM2SI1. Within the first Gyr of our GM3-suite simulations, the GM3 simulation has formed about 2 times as many SMBHs as the GM3SI1 simulation. After 1 Gyr, the GM3 galaxy quickly produces about 10 more SMBHs whereas the GM3SI1 galaxy production flattens out, producing only 1 more SMBH for the remainder of the simulation. The difference between GM2-suite galaxy simulations are less drastic, which is reflected in differences in star formation. 
During the first Gyr, SMBH production is similar until near $
t = 1$~Gyr, where GM2 has produced about 20 SMBHs, compared to GM2SI1 which has produced around 15 SMBHs. After 1 Gyr, the GM2 simulation continues to produce SMBHs, while the GM2SI1 simulation has nearly stopped. 

\begin{figure} 
\centering 
\includegraphics[width=\columnwidth]{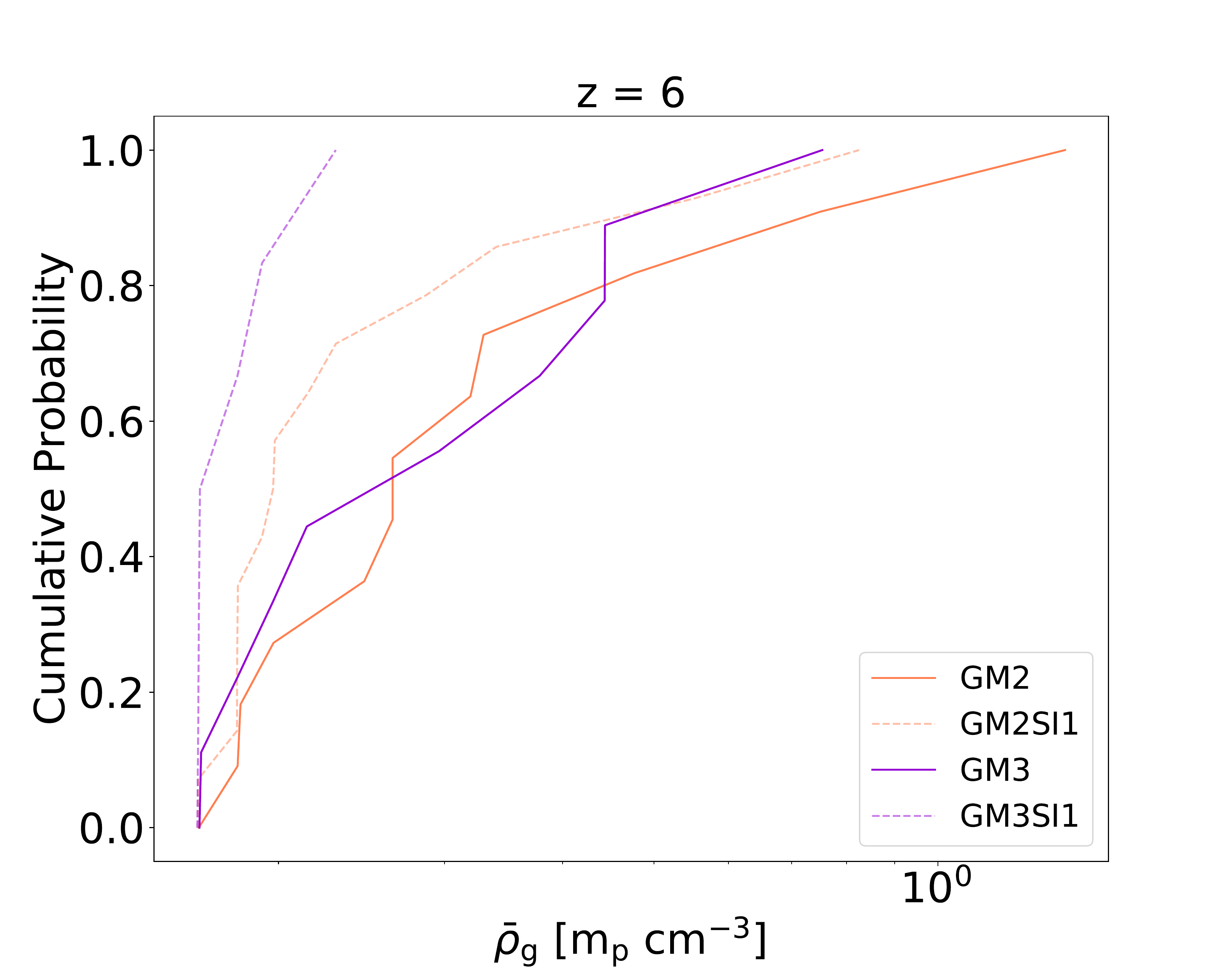} 
    \caption{Cumulative probability vs. average central gas density (gas within 500 pc) for haloes within the mass range for SMBH formation at $z = 6$, the redshift snapshot near ``peak" SMBH production in the simulations. The SIDM galaxies have higher cumulative probability of having lower central gas densities compared to the CDM galaxies.}
    \label{fig:densityprof}
\end{figure}

Our SMBH seeding prescription depends on the gas density in haloes in the high-$z$ Universe. To determine how the halo densities have effected the formation of the SIDM SMBHs, we look at all redshift snapshots before $z = 6$ and determine the approximate halo mass range in which SMBHs form. In the $z = 6$ snapshot, the snapshot closest to peak SMBH production, we then examine the average gas density within 500 pc of the haloes shrink sphere center of the haloes in this mass range. In Figure \ref{fig:densityprof} we plot the cumulative probability to have a given average central gas density vs. average gas density. We find that the SIDM simulations tend to have higher cumulative probabilities at lower average central gas densities compared to the CDM simulations. We conduct a two-sample Kolmogorov-Smirnov test and determine the in the GM2-suite, the null hypothesis that the samples are drawn from the same distribution, is rejected at the 0.24 level. In the GM3-suite the null hypothesis is rejected at the 0.16 level. Further, we find that the DM component dominates the total galaxy density in these haloes, with the baryon density following the DM component \citep[see also][]{Vogelsberger2014}. 

\begin{figure}
\centering 
\includegraphics[width=\columnwidth]{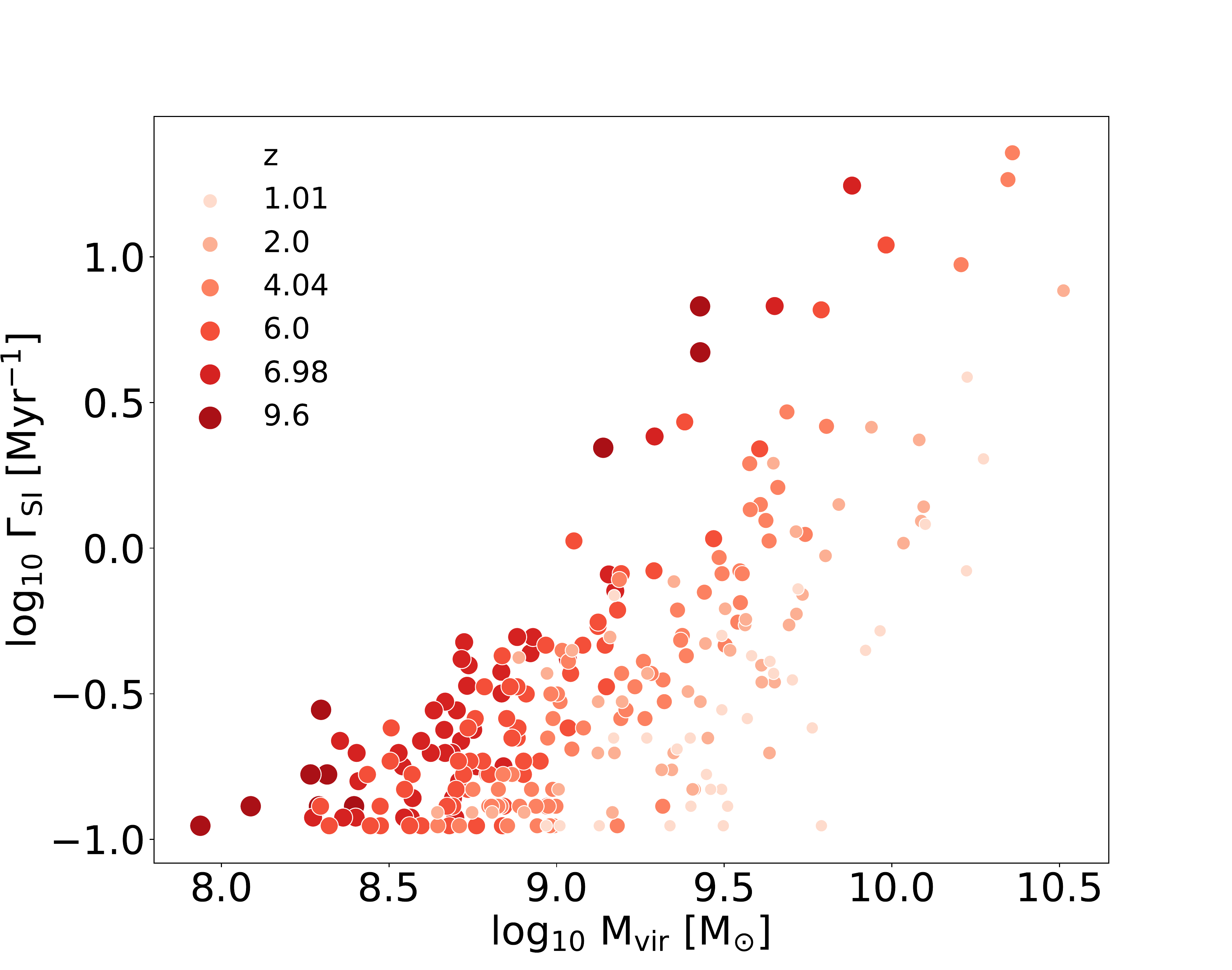} 
    \caption{SIDM interaction rate as a function of halo mass at various redshifts.  Mass range selected to emphasize halo mass range of SMBHS formation sites at  $z \gtrapprox$~6. At a given mass, the SIDM interaction rate is higher at higher redshift, consistent with the fact the high-$z$ universe is denser than the low-$z$ universe.}
    \label{fig:Rate}
\end{figure}

To determine the influence of SIDM on DM, and subsequently gas, densities at high-$z$, we used Lagrangian particle tracing to calculate the average SIDM interaction rate at a given halo mass for a number of redshifts. For each final redshift snapshot, we traced back DM particles to the previous snapshot and calculated the change in the cumulative number of interactions for each halo. We then found the corresponding time interval between the snapshots and calculated the average interaction rate. In Figure \ref{fig:Rate} we plot the total interaction rate vs. halo mass at various redshifts. We find that, at a given halo mass, the average SIDM interaction rate increases towards higher redshift. This trend as well as the general shapes of our interaction rate as a function of $M_{\mathrm{vir}}$ at different redshifts are in agreement with previous analytic work on the cosmic evolution of SIDM interaction rates \citep[see for example the second panel of Figure 1 in ][]{Roberston2015}. Further, at $z = 6$, just before the peak of SMBH formation in the CDM galaxies, the SIDM interaction rate is more than an order of magnitude higher than it is at $z = 1$. We attribute this increase to the increasing mean density of the Universe at higher redshift (see Equation \eqref{eqn:SIrate}). 

Finally, we hypothesize that the decrease in central gas densities and subsequent suppression of SMBH formation is due to the DM component being suppressed in the central region due to DM self-interactions and the baryon density following the DM component. Here we've made several measurements to test this: 1) the SIDM interaction rate appears to be sufficient to relax the cores at high-$z$, and 2) our measurement of the central gas density is consistent with it being lower in the SIDM simulations due to DM coring. 

\begin{figure}
\centering
\vspace{0.5mm}  
\includegraphics[width=\columnwidth]{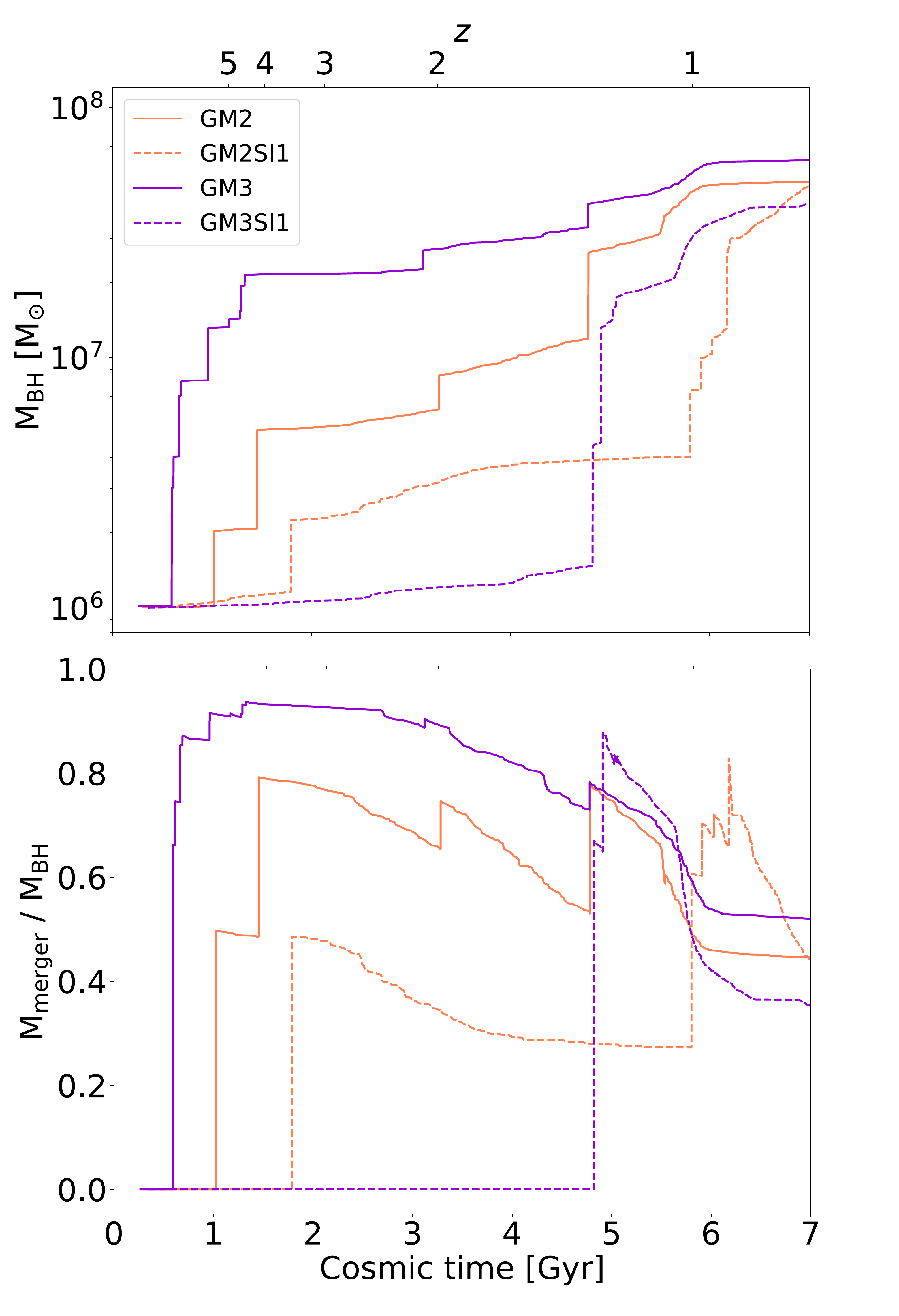}
    \caption{\textit{Top panel}: Mass of the most massive SMBH as a function of cosmic time in each simulation. In both GM2 and GM3 CDM simulations the SMBHs grow more rapidly through mergers before $t = 2$~Gyr. The increased number of SMBHs in the CDM galaxies will lead to more SMBH-SMBH mergers in the early universe, resulting in enhanced SMBH mass growth through mergers in the first 2 Gyrs in the CDM galaxies. This enhanced growth contributes to fast SMBH accretion in the CDM galaxies, given modified Bondi-Hoyle accretion is proportional to $M_{\mathrm{BH}}^2$. \textit{Bottom panel}: BH merger mass divided by $M_{\mathrm{BH}}$ as a function of cosmic time in each simulation. At early times $M_{\mathrm{merger}}$ / $M_{\mathrm{BH}}$ in GM2SI1 and GM3SI1 is 0, indicating delayed merging in SIDM. The sharp jumps in $M_{\mathrm{merger}}$ / $M_{\mathrm{BH}}$ are due to merger events. The smooth decreasing seen is due to smooth accretion in $M_{\mathrm{BH}}$. }
    \label{fig:BHmass}
\end{figure}

\subsection{Black Hole Mergers and Accretion} \label{subsec:BHgrowth}

SMBH-SMBH mergers become more frequent in the early universe in our CDM simulations. This can be seen in the top and bottom panels of Figure~\ref{fig:BHmass}. The top panel shows the growth of the most central SMBH in the largest progenitors of our most massive haloes at $z = 0$. This mass increases due to SMBH-SMBH mergers and through our modified Bondi-Holye accretion prescription detailed in Equation \eqref{eqn:Mdot}. The bottom panel of Figure ~\ref{fig:BHmass} shows $M_{\mathrm{merger}}$, the mass of the SMBH acquired through mergers divided by the total SMBH mass vs. cosmic time. We define the SMBH merger mass to be the total SMBH mass minus the seed mass minus the mass acquired through accretion. This panel clearly shows that SIDM SMBH mergers are delayed by billions of years compared to their CDM counterparts. This merger growth also translates to the sharp jumps present in the top panel of Figure~\ref{fig:BHmass}. Merger rates can be influenced by the total number of SMBHs formed (halo occupation fraction), or by decreased dynamical friction due to SIDM as found in \citet{DiCintio2017b}. Here we demonstrate that the merger rates are correlated with the number of SMBHs that have formed in our simulations, see Figure~\ref{fig:BHhist}. However, decreased dynamical friction also likely further delays SMBH growth in the SIDM galaxies. The main SMBHs of the largest progenitors in our CDM runs subsequently acquire large boosts in SMBH mass from increased merging. The increased growth in the CDM SMBHs due to merging will have large effects on the subsequent evolution of our CDM simulated galaxies, as SMBH accretion is proportional to $M_{BH}^2$. 

\begin{figure}
\includegraphics[width=\columnwidth]{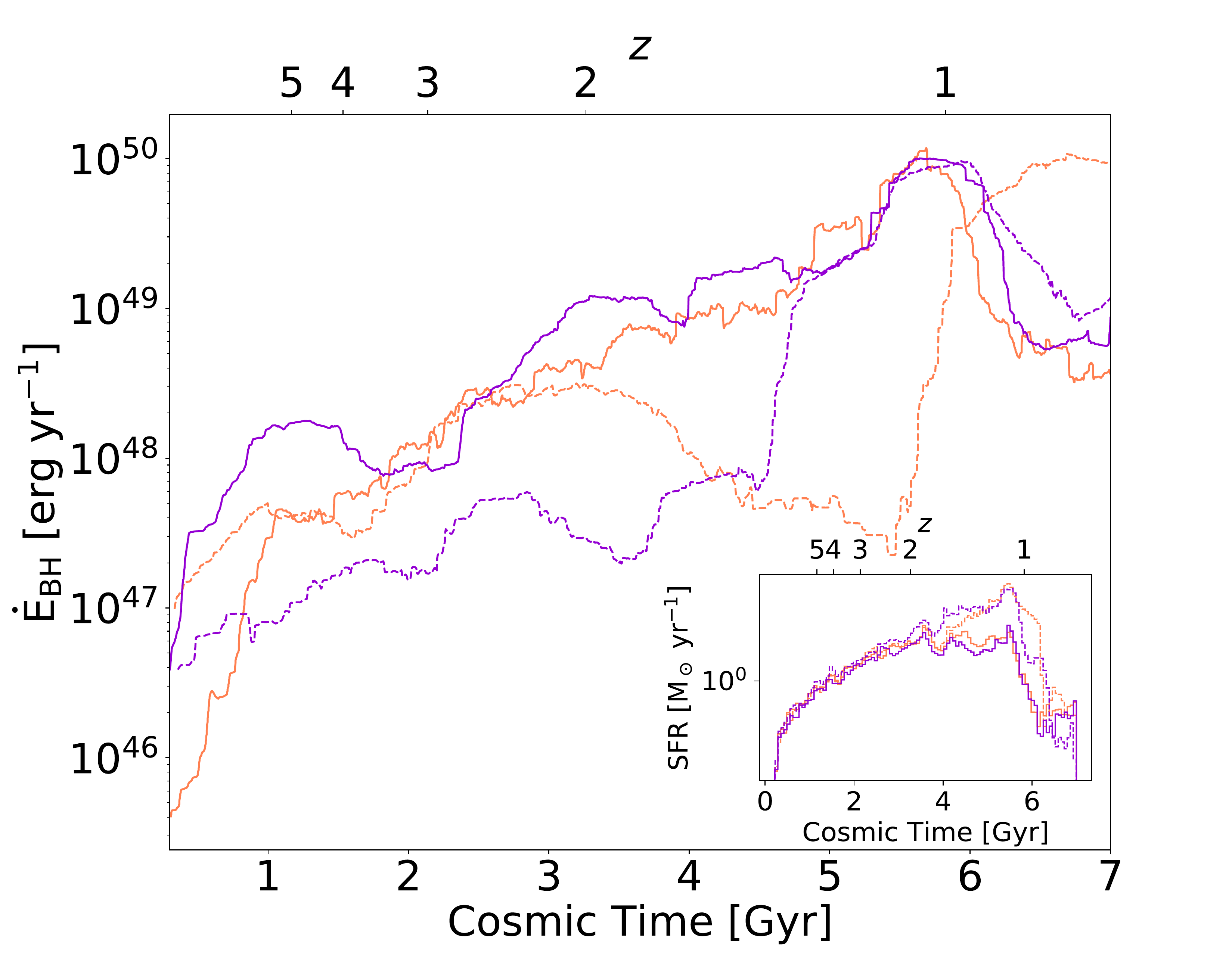}
    \caption{Rate of energy injected from the most massive SMBH ($\dot{E}_{\mathrm{BH}}$) of the major progenitors in each simulation at $z = 0$ vs. cosmic time. Same color-scheme as previous figures. In the GM3-suite there is a large offset between GM3 and GM3SI1 for $t< 4.5$, whereas in the GM2-suite the the injected SMBH energy in GM2 surpasses the GM2SI1 SMBH starting at $t = 3$ Gyr, with the difference growing out to 6 Gyrs. The SMBH energy injected is correlated with the SFR of the host galaxies in all 4 simulations. \textit{Inset}: Log of SFR vs. cosmic time in all 4 simulations to compare with $\dot{E}_{\mathrm{BH}}$}
    \label{fig:BHeninj}
\end{figure}

SMBH accretion has been linked with regulating star formation in MW-mass galaxies \citep{Silk1998,DiMatteo2005,Croton2006} as SMBH accretion produces feedback which can heat up and displace gas from the central regions of galaxies (and large fractions of the star forming disk region of galaxies). The suppressed SMBH-SMBH merger rate in the SIDM galaxies relative to the CDM galaxies causes their central SMBHs to grow through mergers less efficiently. In the Bondi formalism, a lower-mass SMBH accretes less mass than a more massive counterpart, therefore the SMBHs in the SIDM run have suppressed accretion with respect to the CDM runs. The suppressed SMBH accretion further suppresses SMBH feedback, thus regulating star formation less effectively in the SIDM galaxies relative to the CDM galaxies. This results in a larger production of stars in the SIDM galaxies compared to the CDM galaxies. This can be seen by examining Figure~\ref{fig:BHeninj}.

Starting with the GM2-Suite we see that from $t = 0.5-1$~Gyr, the energy injected from the central SMBHs is greater in the SIDM galaxy, compared to the CDM galaxy, however star formation is small compared to the mean SFR during this epoch and thus is unaffected by this  difference. From $t\approx$ 1-3 Gyr the SMBH energy injected is comparable in GM2 and GM2SI1 and the star formation in the two galaxies is very similar. At $t = 3$~Gyr, the energy injected from GM2 starts to increase relative to GM2SI1 and the difference in injected energy continues to grows to $t = 6$~Gyr. A gigayear after the two SMBHs start to deviate in injected SMBH energy, GM2SI1 starts to produce more stars compared to GM2 and continues to produce more stars until both galaxies quench near $t\approx$ 6 Gyr. 

Differences in the GM3-suite are more apparent starting from $t = 0.5$~Gyr. There is a large difference in energy injected from GM3 and GM3SI1 from $t = 0.5-6$~Gyr. These differences can be traced in the SFH. Again, at early times star formation is small compared to the mean SFR, and thus it is not until $t \approx 2.5$ that we start to see deviations in star formation in GM3SI1 compared to GM3. Delayed growth of SMBHs in SIDM relative to CDM and subsequent suppressed accretion results in very different galaxies in SIDM compared to CDM even when you start with identical ICs, as can be seen in Figure \ref{fig:uvi}.

\section{Discussion and Conclusions} \label{sec:discussion}

In this study we used fully cosmological galaxy simulations in CDM and SIDM with a constant cross-section of 1 cm$^2$/g to examine how the co-evolution of SMBHs and their MW-mass host galaxies ($M_{\mathrm{vir}} \approx 7 \times10^{11} M_{\odot}$ at $z = 0$) is influenced by different DM models. To do this we used physically motivated models of SMBH formation and growth (T15; T17) and simulated a MW-mass galaxy with quenched star formation in CDM and SIDM to maximize the effects of SMBH growth on galaxy evolution. We then ran a genetically modified \citep{Roth2016} version of these galaxies to check for result dependent sensitivity to small changes. We found that: 

\indent $\bullet$ SMBH formation is consistently suppressed in SIDM relative to a classic $\Lambda$CDM cosmology. Our CDM simulations produced about 2 or 3 times as many SMBHs compared to our SIDM simulations. \\

\indent $\bullet$ SIDM delays SMBH growth through mergers by billions of years compared to CDM growth. \\
\indent $\bullet$ SIDM SMBHs generate less SMBH feedback compared to CDM SMBHs during the first 5 Gyrs of their evolution. Nonetheless, by $z = 0.8$ their SMBH masses differ only by around 02. dex , so that both CDM and SIDM runs remain compatible with the $M_{BH}-M_{*}$ relation \citep{Schramm2013}.\\
\indent $\bullet$ SIDM galaxies have a larger central reservoir of gas available for star formation. \\
\indent $\bullet$ SIDM galaxies form  about 3 or 4 times more stars than CDM galaxies over their lifetimes. \\

Importantly, \citet{DiCintio2017b} also finds less massive SMBHs and more stars in their MW-mass galaxies from abundance matching at a much higher SIDM cross-section of $\sigma_{\mathrm{dm}} = 10$ cm$^2$/g. At lower masses, the effects of delayed SMBH growth should not matter much given that SMBHs grow very little in dwarf galaxies \citep{Volonteri2008, Habouzit2017, Bellovary2019}. SMBHs are thus not expected to have a significant impact in regulating star formation in dwarf galaxies \citep[but see][]{Silk2017, Sharma2019}. 

It should be noted that, similar to \citet{DiCintio2017b}, these simulations do not have high enough resolution to produce dark matter cores in CDM due to bursty SNe feedback. However this will not significantly alter the results of this study. There is not enough star formation in the CDM galaxies during the epoch of SMBH formation to produce cores since these galaxies have stellar mass less-than or equal to $10^6 M_{\odot}$, the threshold for star formation to start to core DM haloes \citep[see Figure 7 of ][]{Governato2015}. At late times, baryons dominate the central regions of MW-mass galaxies, dramatically shrinking cores formed in SIDM \citep{Dicinto2014a,Kaplinghat2014}. 

Despite the fact that our simulations include well-motivated models of SMBH formation and growth, they are still relatively simple subgrid models. Further, our SMBH subgrid parameters are based on matching observations at $z = 0$ in a $\Lambda$CDM cosmology and thus future work requires an investigation of how and if these parameters might be altered if instead matched to $\Lambda$SIDM. On the other hand, both CDM and SIDM runs remain compatible with the $z = 0.8$ $M_{BH}-M_{*}$ relationship found in \citet{Schramm2013} thus indicating that SIDM does not demand a re-calibration of feedback. Further, in terms of SMBH seeding parameters, most SMBH formation models require high density, very low metallicity gas with similar threshold values to those used in this work \citep[see][]{Volonteri2012}. Future work requires a larger simulation with higher output resolution to more thoroughly quantify how SIDM influences gas densities at high-$z$.  However, based on the tests conducted in this work, our hypothesis that SIDM lowers central gas densities due to self-interactions at high-$z$ holds.  Thus, given our limitations and small sample size, this study is a useful case study which shows that, given a well-motivated SMBH formation prescription, SIDM can significantly alter SMBH merger histories and delay growth and feedback which results in very different galaxy evolution of MW-mass objects compared to the classic $\Lambda$CDM model.

\section*{Acknowledgements}

For our simulation analysis we utilized \textsc{pynbody} \citep{Pontzen2013}  and \textsc{tangos} \citep{Pontzen2018}. This material is based upon work supported by the National Science Foundation Graduate Research Fellowship under Grant No. DGE-1762114. AC would like to thank Matt McQuinn, Vid Irsic, Jessica Werk, Annika Peter, Charlotte Christensen, Iryna Butsky, Jillian Belovary, Ray Sharma, and Sarah Loebman for useful conversations through out the progression of this work. ADC acknowledges financial support from a Marie-Sklodowska-Curie Individual Fellowship grant, H2020-MSCA-IF-2016 Grant agreement 748213, DIGESTIVO. AP was supported by the Royal Society. Resources supporting this work were provided by the NASA High-End Computing (HEC) Program through the NASA Advanced Supercomputing (NAS) Division at Ames Research Center. This work used the Extreme Science and Engineering Discovery Environment (XSEDE), which is supported by NSF grant number ACI-1548562. This work has received funding from the European Union’s Horizon 2020 research and innovation programme under grant agreement No. 818085 GMGalaxies.












\bsp	
\label{lastpage}
\end{document}